\documentclass[aps,twocolumn,longbibliography,notitle] {revtex4-1}
\usepackage{amsmath,graphicx,amsfonts}
\usepackage{subcaption}
\usepackage{times}
\usepackage[colorlinks=true,linkcolor=blue,urlcolor=blue,citecolor=blue, breaklinks=true,hypertexnames=false]{hyperref}
\usepackage{xcolor}
\usepackage{amssymb}
\usepackage{bbm}
\usepackage{physics}
\usepackage{braket}
\usepackage{comment}
\usepackage{placeins}
\usepackage[utf8]{inputenc}

\DeclareUnicodeCharacter{03BC}{\mbox{$\mu$}}

\DeclareUnicodeCharacter{03B4}{$\delta$}
\DeclareUnicodeCharacter{2009}{-}

\newcommand{\lcpq}{Univ Toulouse, CNRS, LCPQ, Toulouse, France}
\newcommand{\lpt}{Univ Toulouse, CNRS, LPT, Toulouse, France}
\newcommand{\etsf}{European Theoretical Spectroscopy Facility (ETSF), Toulouse, France}

\usepackage[english]{babel}
\addto\captionsenglish{}
\addto\captionsenglish{}

\makeatletter
\def\bbl@set@language#1{%
    \edef\languagename{%
        \ifnum\escapechar=\expandafter`\string#1\@empty
        \else\string#1\@empty\fi}%
    \@ifundefined{babel@language@alias@\languagename}{}{%
        \edef\languagename{\@nameuse{babel@language@alias@\languagename}}%
    }%
    \select@language{\languagename}%
    \expandafter\ifx\csname date\languagename\endcsname\relax\else
    \if@filesw
    \protected@write\@auxout{}{\string\select@language{\languagename}}%
    \bbl@for\bbl@tempa\BabelContentsFiles{%
        \addtocontents{\bbl@tempa}{\xstring\select@language{\languagename}}}%
    \bbl@usehooks{write}{}%
    \fi
    \fi}
\newcommand{\DeclareLanguageAlias}[2]{%
    \global\@namedef{babel@language@alias@#1}{#2}%
}
\makeatother

\DeclareLanguageAlias{en}{english}

\makeatletter
\def\@bibdataout@aps{%
    \immediate\write\@bibdataout{%
        @CONTROL{%
            apsrev41Control%
            \longbibliography@sw{%
                ,author="08",editor="1",pages="1",title="0",year="1"%
            }{%
                ,author="08",editor="1",pages="1",title="",year="1"%
            }%
        }%
    }%
    \if@filesw \immediate \write \@auxout {\string \citation {apsrev41Control}}\fi
}
\makeatother

\date\today

\def\TITLE{Electronic and magnetic properties of small one-dimensional Wigner crystals from an \textit{ab initio} approach}

\begin{document}
\raggedbottom
\title{\TITLE}

\author{Daniele Lagasco}
\email{daniele.lagasco@irsamc.ups-tlse.fr}
\affiliation{\lpt}
\affiliation{\etsf}
\author{J. Arjan Berger}
\email{arjan.berger@irsamc.ups-tlse.fr}
\affiliation{\lcpq}
\affiliation{\etsf}


\begin{abstract}
    We present an \emph{ab initio} method to study the electronic properties of small one-dimensional Wigner crystals. In particular, we focus on the calculation of the electronic charge distribution. Our theoretical studies are motivated by the experimental observation of few-electron Wigner crystals in a carbon nanotube [Science 364, 870 (2019)]. We model the experimental setup by confining electrons in a one-dimensional potential well with hard-wall boundary conditions. We represent the Hamiltonian of the system in a basis of Slater determinants and perform full configuration interaction to ensure we capture all the electron correlation for a given basis set. As the one-particle basis set we use particle-in-a-box wave functions which by construction satisfy the boundary conditions. With our approach, we obtain accurate electronic density profiles of small one-dimensional Wigner crystals. These profiles clearly show the localisation of the electrons. Moreover, we calculate the exchange coupling constant of the Wigner dimer by mapping our \textit{ab initio} method on a Heisenberg Hamiltonian. We obtain a reasonable agreement with a result in the literature.
\end{abstract}

\maketitle

 \section{Introduction}
\label{Introduction}
Wigner crystallization, predicted by Eugene Wigner in 1934 \cite{Wigner34}, is a hypothesized phase of matter in the electronic liquid in which the electrons are localized. This phase is expected to exist at low density, when the Coulomb interaction energy dominates over the kinetic one. The experimental observation of Wigner crystallization has been elusive for a long time since at low densities the appearance of disorder and invasive measurement techniques destroys the crystalline order. The earliest indications of Wigner crystallization were obtained using the high magnetic-field
transport measurements in semiconducting two-dimensional systems \cite{2Dsemicond_1,2Dsemicond_2} as well as in a sheet of electrons on the surface of liquid helium \cite{LiquidHe}. Very recently, the direct observation of Wigner crystals has been achieved \cite{Experiment,WCmonolayer,TMDHetWC,GrapheneWC}.

In this work we study the electronic and magnetic properties of one-dimensional (1D) Wigner crystals. Our theoretical investigations are motivated by the experiment performed by I. Shapir and coworkers \cite{Experiment}. In that work, the authors managed to control the injection of few electrons (from 2 to 6) in a carbon nanotube  of length $L=1600\, \mathrm{nm}$, which can be considered a 1D system. They probed and imaged the spatial distribution of the electrons in the nanotube and observed that the electrons are localized.

From the theoretical side, Wigner crystallization has been studied by using effective phenomenological models \cite{effHam,pot_rin,Bosonization_Glazman,Bosonization_Schulz,Wigner_oscT2014,Wigner_locT2016,Wigner_corrT2013,review_WC}  or by \textit{ab initio} non-relativistic Hamiltonians. In the latter case, many techniques have been employed to extract properties of few-electron Wigner crystals, such as exact diagonalization~\cite{jauregui1993,FCI_Rontani,ED_Roy_2009,ED_Roy_2012,ED_Rontani_2010,FCI_gauss,ED_Berger,Ab_initioWC_Berger}, density functional theory~\cite{DFT}, Hartree-Fock~\cite{Escobar2024} and semiclassical methods~\cite{semicl_pot_Berger,mesoscopic_eff,Wunsch2009,Hausler_WKB,Matveev_J,J_WKB,rings}. Recently, the exchange coupling of a Wigner dimer was studied by solving analytically the related Schr\"odinger equation and by using the semi-classical WKB method \cite{ex_dimer}. 

In this work, we model the experimental system with a Hamiltonian describing $N$ interacting electrons in a 1D well of length $L$ with impenetrable side barriers. 
We then calculate the ground- and excited-state wave functions using full configuration interaction.
This \textit{ab initio} approach ensures that we capture all the electron correlation for a given basis set.
This is crucial, since a Wigner crystal is a strongly correlated system. 
From the ground-state wave function, we can then calculate the ground-state electron density.
We then obtain the eigenvalues and eigenvectors, i.e., the ground and excited state wave functions, of the Hamiltonian.
We will show that the density profiles of small 1D Wigner crystals with $N$ electrons show $N$ distinct peaks as expected.
This shows that the electrons are localized because of the strong electronic repulsion.

It is known that a 1D Wigner crystal has an antiferromagnetic spin order \cite{antiferro,Ferro1D}. 
Therefore, its low-energy properties can be mapped onto an antiferromagnetic Heisenberg-Dirac-Van Vleck (HDVV) Hamiltonian \cite{HDVV_heisenberg,HDVV_dirac,HDVV_van_vleck}, which describes a system of localized interacting electrons. This model depends on magnetic exchange coupling constants, which are the parameters that control the interactions among the electrons.  In this work, we calculate the exchange coupling $J$ of a Wigner dimer.

The remainder of the article is organised as follows:
in Sec.~\ref{theory} we present the theory of our approach while in Sec.~\ref{results} we report the results for the density profiles of the 1D Wigner crystals and the exchange parameter of the Wigner dimer.
Finally, in Sec.~\ref{conclusion} we give the conclusions and prospects of this work.
\section{Theory}
\label{theory}
\subsection{Numerical method for the study of small one-dimensional Wigner crystals}
\label{method}
As mentioned in the introduction, we set out to model electrons that are confined to a nanotube of length $L=1600 \, \mathrm{nm}$ and radius $R=0.5 \, \mathrm{nm}$ from first principles. Since $\frac{R}{L} \ll 1$ we model this system using a 1D many-body Hamiltonian with hard-wall boundary conditions. It is given by
\begin{equation}
    \hat{H}=-\frac{\hbar^2}{2m_{\text{eff}}}\sum_{i}^{N}\frac{\partial^2}{\partial z^2_i}+\frac{e^2}{2}\sum_{i\not = j}V(z_i-z_j)\,, \quad z_i\in[0,L]
    \label{Ham}
\end{equation}
where $z_i$ is the coordinate of the $i^{th}$ electron, $m_{\text{eff}}$ is the effective mass of the electron, which is given by $m_{\text{eff}}=0.0062 m_e$, which was obtained from the experimental band structure~\cite{Experiment}.
The potential $V(z_i-z_j)$ accounts for the Coulomb repulsion between two electrons with coordinates $z_i$ and $z_j$.

In one dimension, the standard Coulomb potential $1/|z_i-z_j|$ has a non integrable singularity at  ($z_i=z_j$). Therefore, instead, we employ the following regularized Coulomb potential proposed by Sarkany \textit{et al.}~\cite{pot_rin} for electrons confined to nanotubes of radius $R$,
\begin{multline}
     V(z_i-z_j)=\frac{2}{\pi}\frac{1}{\sqrt{(z_i-z_j)^2+4R^2}}\\ K\Bigg(\frac{2R}{\sqrt{(z_i-z_j)^2+4R^2}}\Bigg)\,
     \label{rin_Coul_pot},
\end{multline}
where $K(x)$ is the complete elliptic integral of the first kind~\cite{special_func}.
The interaction in Eq.~\eqref{rin_Coul_pot} describes the electronic screening felt by electrons in a carbon nanotube of radius $R$~\cite{pot_rin}. For more details see Ref.~\cite{pot_rin}.

It is instructive to investigate the asymptotic expressions of the regularized Coulomb potential for 
$\frac{|z_i-z_j|}{R}\gg1 $ and $\frac{|z_i-z_j|}{R}\ll1 $.
They are given by
\begin{align}
    &V(|z_i-z_j|)= \notag \\
    &\left\{
    \begin{aligned}
    &\frac{1}{|z_i-z_j|}\qquad && \text{if}\quad \frac{|z_i-z_j|}{R}\gg1\,, \\
    &\frac{1}{\pi R}\big[\ln{\sqrt{2}} +\ln{(8R/|z_i-z_j|)} \big]\quad && \text{if} \quad \frac{|z_i-z_j|}{R}\ll1\,.
    \end{aligned}
\right.
\label{asymptoticV}
\end{align}
We observe that if the electrons are well separated, i.e., $\frac{|z_i-z_j|}{R}\gg1$, the regularized Coulomb potential tends to the standard Coulomb potential as it should. Instead, if the electrons are close to each other, i.e., $|z_i-z_j|/R\ll1$, the regularized Coulomb potential has an integrable singularity.

To simplify the Hamiltonian, we rescale the Cartesian coordinates according to $x_i=z_i/L$ and we introduce the effective Bohr radius $a_B=\frac{\hbar^2}{m_{\text{eff}}e^2}=8.5\, \mathrm{nm}$.
We can thus express the Hamiltonian in Eq. \eqref{Ham} as
\begin{align}
    \hat{H}=\frac{\hbar^2}{2m_{\text{eff}} L^2}\Bigg[&-\sum_{i=1}^{N}\frac{d^2}{dx_i^2}+2\frac{L}{a_B}\sum_{i<j}^{N}V(x_i-x_j)\Bigg]\,, \nonumber \\
    & x_i\in[0,1]\label{adham}\, ,
\end{align}
where  $\frac{L}{a_B} = 188.235$.
We note that the rescaling of the coordinates also leads to a rescaling of the radius $R$ in Eq.~\eqref{adham} according to $R/L$.
Due to the hard-wall boundary conditions, we have to solve for the eigenvalues and eigenfunctions of the Hamiltonian in Eq.~\eqref{adham} with the boundary condition that the eigenfunctions, i.e., the ground- and excited-state wave functions, have to vanish if at least one of the coordinates $x_1,...,x_N$ is $\le 0$ or $\ge 1$. 

Since we are in the strongly correlated regime we will use the method of full configuration interaction (FCI) to guarantee we capture all electron correlation for a given basis set.
Therefore, we write the eigenfunctions of $\hat{H}$ as linear combinations of Slater determinants $\Phi_{\alpha}$. 
Their general expression is
\begin{equation}
    \Psi_n(y_1,...,y_N)=\sum_{K}c_{nK}\Phi_{K}(y_1,...,y_N)\, ,
    \label{MBeigfs}
\end{equation}
where $y_i=(x_i,\omega_i)$ with $\omega_i$ the spin coordinate of the spin function $\sigma(\omega_i)$ of electron $i$ which can be either $\alpha(\omega_i)$ (spin up) or $\beta(\omega_i)$ (spin down). The Slater determinants are given by
%
%
\begin{equation}
\label{Sldet_spin}
\Phi_{K}(y_1,y_2,\ldots,y_N)
=
\frac{1}{\sqrt{N!}}
\begin{vmatrix}
\psi_m(y_1) & \psi_n(y_1) & \cdots & \psi_p(y_1) \\
\psi_m(y_2) & \psi_n(y_2) & \cdots & \psi_p(y_2) \\
\vdots & \vdots & \ddots & \vdots \\
\psi_m(y_N) & \psi_n(y_N) & \cdots & \psi_p(y_N)
\end{vmatrix},
\end{equation}
in which $\psi_n(y) = \chi_n(x)\sigma_n(\omega)$ are the spinorbitals with $\chi_n(x)$ the one-electron basis functions.

We choose $\chi_n$ to be particle-in-a-box wave functions given by
\begin{align}
     \chi_n(x)&=\sqrt{2}\sin(n \pi x)\,, \quad n \in \mathbb{N}
     \nonumber \\  \chi_n(x) &= 0 \quad \text{if $x<0 \vee x>1$}\,.
    \label{eigf}
\end{align}
This choice ensures that the boundary conditions imposed on the many-body wave functions are automatically satisfied.

In the FCI approach we have to evaluate the matrix elements of $\hat{H}$ in terms of the Slater determinants.
Since the many-body Hamiltonian is a two-body operator, 
it is convenient to split these matrix elements $\langle \Phi_{K}|\hat{H}|\Phi_{L}\rangle$ into four classes according to the Slater Condon rules (see, e.g., \cite{Szabo}),
\begin{enumerate}
	\item $\ket{\Phi_{K}} = \ket{\Phi_{L}}$.
	\item $\ket{\Phi_{K}}$ and $\ket{\Phi_{L}}$ only differ in spinorbitals $\psi_m$ and $\psi_p$.
	\item $\ket{\Phi_{K}}$ and $\ket{\Phi_{L}}$ only differ in the couple of spinorbitals $\{\psi_m, \psi_n\}$ and $\{\psi_p, \psi_q\}$.
    \item $\ket{\Phi_{K}}$ and $\ket{\Phi_{L}}$ differ by more than two spinorbitals.
\end{enumerate}
The general expressions of $\langle \Phi_{K}|\hat{H}|\Phi_{L}\rangle$ for the four cases are given by the following relations 
\begin{enumerate}
    \item 
    \begin{align}
	\bra{\Phi_K}\hat{H}\ket{\Phi_K}&=
    \frac12\bigg(\sum_{\psi_m \in \Phi_K}\sum_{\psi_n \in \Phi_K} \!\! A[m,n,m,n]\delta_{\sigma_m,\sigma_m}\delta_{\sigma_n,\sigma_n}
    \nonumber \\ & -
    A[m,n,n,m]\delta_{\sigma_m,\sigma_n}\delta_{\sigma_m,\sigma_n}\bigg) + \pi^2 \sum_{\psi_n\in\Phi_K} \!\! n^2 ,
    \label{form1spin}
    \end{align}
	\item
    \begin{align}
    \bra{\Phi_K}\hat{H}\ket{\Phi_L}&={(-1)}^Q
    \!\! \sum_{\psi_n\in \Phi_K \cap \Phi_L} \!\! \big[ A[m,n,p,n]\delta_{\sigma_m,\sigma_p}\delta_{\sigma_n,\sigma_n}
    \nonumber \\ & -
    A[m,n,n,p]\delta_{\sigma_m,\sigma_n}\delta_{\sigma_n,\sigma_p}\big]\,,\label{form2spin}
    \end{align}
	\item 
    \begin{align}
    \bra{\Phi_K}\hat{H}\ket{\Phi_L}&={(-1)}^Q[ A[m,n,p,q]\delta_{\sigma_m,\sigma_p}\delta_{\sigma_n,\sigma_q}
    \nonumber \\ & -
    A[m,n,q,p]\delta_{\sigma_m,\sigma_q}\delta_{\sigma_n,\sigma_p}\big]\,\label{form3spin}\, ,
    \end{align}
    \item 
    \begin{equation}
    \bra{\Phi_K}\hat{H}\ket{\Phi_L} = 0 \label{form4spin}\, ,
    \end{equation}
\end{enumerate}
where $Q$ corresponds to the number of permutations required to obtain the maximum coincidence between the Slater determinants $\ket{\Phi_K}$ and $\ket{\Psi_L}$~\cite{Szabo} and
\begin{align}
A[m,n,p,q] &= 2\frac{L}{a_B}\int_{0}^{1}\,dx_1\int_{0}^{1}dx_2 V(x_1-x_2) 
\nonumber \\ &\times
\chi_m(x_1)\chi_n(x_2)\chi_p(x_1)\chi_q(x_2).
\label{Aijkl1}
\end{align}
We note that the contribution from the kinetic operator to $\langle \Phi_{K}|\hat{H}|\Phi_{L}\rangle$ vanishes whenever $\ket{\Phi_K} \neq \ket{\Phi_L}$ because of the orthogonality of the one-electron basis functions.

\subsection{Spin adaption of the basis}
\label{spin_adaptation}
Each electronic wave function has a certain spin symmetry, and wave functions with different spin symmetries do not mix, i.e., their overlap vanishes.
However, in practice, when the eigenvalues corresponding to two or more wave functions become nearly degenerate, mixing can occur due to the finite precision of numerical calculations.
It is known that in the strong-interaction limit the wave functions corresponding to a given Hamiltonian become degenerate.
Therefore, unphysical mixing of wave functions with different spin symmetries can occur in calculations of Wigner crystals.

To avoid such spin-symmetry mixing, one can perform a spin adaptation of the basis functions.
Within this spin-adapted basis, the eigendecomposition of the full Hamiltonian thus decouples into several smaller eigendecompositions, one for each spin symmetry.
Another important advantage of the spin adaptation is that solving an eigenvalue problem for each spin symmetry separately is numerically much more efficient 
than solving the full eigenvalue problem.
A spin adaptation of the Slater determinants will thus allow us to use large basis sets.

As an illustration, let us focus on the spin adaptation of the Slater determinants describing a Wigner crystal with two electrons.
For this case, the spin symmetries of the wave functions correspond either to spin singlets or spin triplets.  
The singlet wavefunctions are symmetric in the exchange of the spatial coordinates and anti-symmetric with respect to the exchange of spin coordinates and vice versa for the triplet wavefunctions. 
In the singlet sector, we define the spin adapted basis functions $\Psi^{S}_{m,n}$ as
\begin{align}
    &\Psi^{S}_{m,n}(y_1,y_2)=\frac{1}{2\sqrt{(1+\delta_{m,n})}}\big(\chi_m(x_1)\chi_n(x_2)
    \nonumber \\ & +
    \chi_n(x_1)\chi_m(x_2)\big)
    \big(\alpha(\omega_1)\beta(\omega_2)-\beta(\omega_1)\alpha(\omega_2)\big)
    \label{PsiS}
\end{align}
while, in the triplet sector, we define the spin adapted basis functions $\Psi^{T}_{m,n}$ as
\begin{align}
     \Psi^{T}_{m,n}(y_1,y_2)&=\frac{1}{2}\big(\chi_m(x_1)\chi_n(x_2)-\chi_n(x_1)\chi_m(x_2)\big)
     \nonumber \\ & \times 
     \big(\alpha(\omega_1)\beta(\omega_2)+\beta(\omega_1)\alpha(\omega_2)\big)
     \label{PsiT}.
\end{align}
%

A general matrix element of the Hamiltonian in Eq.~\eqref{adham} in the singlet sector is given by
\begin{align}
    & \bra{\Psi^S_{m,n}}\hat{H}\ket{\Psi^S_{p,q}} = (m^2 + n^2) \pi^2\delta_{m,p}\delta_{n,q} +
    \nonumber \\ &  \frac{1}{\sqrt{(1+\delta_{m,n})(1+\delta_{p,q})}}  \Big(A[m,n,p,q] + A[m,n,q,p]\Big)\,.
    \label{singlet_matel}
\end{align}
%
Instead, a general matrix element of the Hamiltonian in Eq.~\eqref{adham} in the triplet sector is given by
\begin{align}
    \bra{\Psi^{T}_{m,n}}\hat{H}\ket{\Psi^T_{p,q}}&=(  m^2 + n^2  ) \pi^2\delta_{m,p}\delta_{n,q}
    \nonumber \\ &+
    A[m,n,p,q]-A[m,n,q,p].
    \label{triplet_matel}
\end{align}

Moreover, we can further reduce the size of the Hamiltonian expressed in the symmetry adapted basis by using the inversion symmetry of the Hamiltonian and the basis functions.
This symmetry guarantees that $A[m,n,p,q]$ vanishes when $m+n$ is even and $p+q$ is odd and when $m+n$ is odd and $p+q$ is even.
The details of this approach can be found in Appendix \ref{Optimized_basis}.
In Appendix \ref{Optimized_basis} we also develop an approach for two electrons that \textit{a priori} selects the Slater determinants with the largest coefficients. This is very useful to rapidly converge the calculations with respect to the number of Slater determinants.

\subsection{The density distribution}
\label{dens_distr}
The electronic density distribution is a quantity that allows us to study the localization of the electrons in the nanotube. 
The density distribution of the ground-state $\rho_0 (x)$ can be obtained from the ground-state wave function $\Psi_0$ according to
\begin{equation}
   \rho_0(x)= \bra{\Psi_0}\hat{\rho}(x)\ket{\Psi_0}\,,
    \label{def_rho}
\end{equation}
where  $\hat{\rho}(x)$ is the density operator, which is defined as
\begin{equation}
    \hat{\rho}(x)=\sum_{i=1}^{N}\delta(x-x_i)\,.
\end{equation}
Using the expansion in Eq.~\eqref{MBeigfs} the density distribution can be rewritten as
\begin{align}
\rho_0(x) &=
\sum_{K} c_{0K}^2 \bra{\Phi_K}\hat{\rho}(x)\ket{\Phi_K} 
\nonumber \\&+ \sum_{K \ne L}c_{0K} c_{0L}\langle \Phi_K|\hat{\rho}(x)|\Phi_L \rangle.
\end{align}

Since the density operator is a one-body operator we can use the Slater-Condon rules for one-body operators to find its matrix elements~\cite{Szabo}.
We have the following three classes
\begin{enumerate}
	\item $\ket{\Phi_{K}} = \ket{\Phi_{L}}$,
	\item $\ket{\Phi_{K}}$ and $\ket{\Phi_{L}}$ only differ in spinorbitals $\psi_m$ and $\psi_p$,
    \item $\ket{\Phi_{K}}$ and $\ket{\Phi_{L}}$ differ by more than one spinorbital.
\end{enumerate}
The general expressions for the matrix elements $\bra{\Phi_K}\hat{\rho}(x)\ket{\Phi_L}$ 
for these three cases are given by
\begin{enumerate}
\item 
    \begin{equation}
	\bra{\Phi_K}\hat{\rho}(x)\ket{\Phi_K}= \sum_{\psi_n \in \Phi_K} \chi^2_n(x),
    \label{form1dens}
    \end{equation}
\item
    \begin{equation}
	\bra{\Phi_K}\hat{\rho}(x)\ket{\Phi_L}=(-1)^Q \chi_m(x)\chi_p(x)\delta_{\sigma_m,\sigma_p},\label{form2dens}
    \end{equation}
\item
    \begin{equation}
    \bra{\Phi_K}\hat{\rho}(x)\ket{\Phi_L} = 0 \label{form3dens},
    \end{equation}
\end{enumerate}
where, as in Sec. \ref{method}, $Q$ corresponds to the number of permutations required to have the maximum coincidence between the Slater determinants $\ket{\Phi_K}$ and $\ket{\Phi_L}$.
\subsection{The 2-body reduced density matrix}
\label{Gamma}
A stronger diagnostic than the electronic density to investigate electron localization is the two-body reduced density matrix. Its diagonal, known as the pair density, describes the joint probability density of finding two electrons at positions $x_1$ and $x_2$.

The two-body reduced density matrix $\Gamma^{(2)}$ is defined as
\begin{align}
&\Gamma^{(2)}(x_1,x_2;x_1^{\prime},x_2^{\prime}) = \frac{N(N-1)}{2} \int d\omega_1\, d\omega_2
\nonumber \\ &
\int dy_3\, ...\,dy_N \Psi^{*}_0(y^{\prime}_1,y^{\prime}_2,y_3,...,y_N)\Psi_0(y_1,y_2,y_3,...,y_N)
\end{align}
Expanding the ground-state wave functions in terms of Slater determinants according to Eq.~\eqref{MBeigfs}, the pair density, which is defined as $\Gamma(x_1,x_2)=\Gamma^{(2)} (x_1,x_2;x_1,x_2)$, can be rewritten as
\begin{equation}
    \Gamma(x_1,x_2) =\frac{N(N-1)}{2} \sum_{K,L} c_{0,K}c_{0,L} I_{K,L}(x_1,x_2)
    \label{pair_corr}
\end{equation}
where
\begin{align}
    I_{K,L}(x_1,x_2) &= \int d\omega_1 d\omega_2 \int dy_3 \cdots dy_N 
   \nonumber \\ & \times 
   \Phi^*_K(y_1,y_2,y_3, \cdots, y_N)\Phi_L(y_1,y_2,y_3, \cdots, y_N),
\end{align}
The integrals $I_{K,L}(x_1,x_2)$ can be obtained using the Slater-Condon rules introduced in Section \ref{method}. The contributions to $\Gamma(x_1,x_2)$ for the various cases are given by
\begin{enumerate}
\item 
    \begin{align}
	&I_{K,K}(x_1,x_2) = \frac{1}{2}\sum_{\chi_m,\chi_n\in \Phi_K}
    \Big[\chi_m(x_1)\chi_n(x_2)\chi_m(x_1)\chi_n(x_2)
    \nonumber \\ \times
    &\delta_{\sigma_m,\sigma_m}\delta_{\sigma_n,\sigma_n}-\chi_m(x_1)\chi_n(x_2)\chi_n(x_1)\chi_m(x_2)\delta_{\sigma_m,\sigma_n}\delta_{\sigma_m,\sigma_n}\Big]
    \label{form1dens}
    \end{align}
\item
    \begin{align}
	&I_{K,L}(x_1,x_2) = \frac{(-1)^Q}{2}\!\!\!\!\!\!\!\!\sum_{\chi_n\in\Phi_K \cap \Phi_L}\!\!
    \Bigg[\chi_m(x_1)\chi_n(x_2)\chi_p(x_1)\chi_n(x_2)
    \nonumber\\ \times
    &\delta_{\sigma_m,\sigma_p}\delta_{\sigma_n,\sigma_n}-\chi_m(x_1)\chi_n(x_2)\chi_n(x_1)\chi_p(x_2) \delta_{\sigma_m,\sigma_n}\delta_{\sigma_n,\sigma_p}\Bigg]
    ,\label{form2dens}
    \end{align}
\item
    \begin{align}
	&I_{K,L}(x_1,x_2) =\frac{(-1)^Q}{2}\big[\chi_m(x_1)\chi_n(x_2)\chi_p(x_1)\chi_q(x_2) \nonumber\\ \times
    &\delta_{\sigma_m,\sigma_p}\delta_{\sigma_n,\sigma_q}-\chi_m(x_1)\chi_n(x_2)\chi_q(x_1)\chi_p(x_2) \delta_{\sigma_m,\sigma_q}\delta_{\sigma_n,\sigma_p}\big]
    \label{form3dens}
    \end{align}
\item $I_{K,L}(x_1,x_2) = 0$
\end{enumerate}

We note that, in the case of two electrons, the pair density is simply given by 
\begin{equation}
\Gamma(x_1,x_2)= \int d\omega_1 d\omega_2 |\Psi_0(y_1,y_2)|^2. 
\end{equation}
The ground-state wave function can then be written as a linear combination of singlet wave functions according to
\begin{equation}
    \Psi_0(y_1,y_2)=\sum_{m,n} c_{0;m,n} \Phi^{S}_{m,n}(y_1,y_2)\,,
\end{equation}
%
%
and the pair density can be calculated as
\begin{align}
    \Gamma(x_1,x_2)&=\sum_{m,n}\sum_{p,q} \frac{c_{0;m,n}c_{0;p,q}}{\sqrt{(1+\delta_{m,n})(1+\delta_{p,q})}} 
    \notag \\ & \times
    [\chi_m(x_1)\chi_n(x_2)+\chi_n(x_1)\chi_m(x_2)]
    \notag \\ & \times
    [\chi_p(x_1)\chi_q(x_2)+\chi_q(x_1)\chi_p(x_2)].
\end{align}
\subsection{The magnetic exchange coupling}
\label{ab_initio_calc}
The low energy properties of a system of localized interacting spins can be captured by the Heisenberg-Dirac-Van Vleck (HDVV) model~\cite{HDVV_heisenberg,HDVV_dirac,HDVV_van_vleck}. The HDVV Hamiltonian is given by
	\begin{equation}
		\hat{H}_{HDVV}=\sum_{i<j}J_{i,j}\hat{S}_i\cdot\hat{S}_j\,,
		\label{eff_Ham}
	\end{equation}
where $\hat{S}_i=\big(\frac{\hbar}{2}\hat{\sigma}_x, \frac{\hbar}{2}\hat{\sigma}_y,\frac{\hbar}{2}\hat{\sigma}_z\big)$, is a generalized vector that represent the localized electrons. This operator contains the Pauli matrices, which are
	\begin{equation}
		\hat{\sigma}_x=
		\begin{pmatrix}
			0 & 1 \\
			1 & 0
		\end{pmatrix}\,,\quad
		\hat{\sigma}_y=
		\begin{pmatrix}
			0 & -i \\
			i & 0
		\end{pmatrix}\,,\quad
		\hat{\sigma}_z=
		\begin{pmatrix}
			1 & 0 \\
			0 & -1
		\end{pmatrix}\,.
    \end{equation}
The parameters $J_{i,j}$ are the exchange coupling constants between the localized electrons. 
We note that $J_{i,j}<0$ implies a ferromagnetic coupling between the spins $S_i$ and $S_J$, while  $J_{i,j}>0$ implies an anti-ferromagnetic coupling between the spins $S_i$ and $S_j$.


Both the Hamiltonian in Eq.~\eqref{adham} and the HDVV Hamiltonian in Eq.~\eqref{eff_Ham}
commute with $\hat{S}^2$ where $\hat{S}$ is the spin angular momentum operator.
Therefore, the eigenvectors of $\hat{H}$ and $\hat{H}_{HDVV}$ have the same spin symmetry.
By comparing the energy differences of two spin states obtained from diagonalizing $\hat{H}$ and $\hat{H}_{HDVV}$ we can obtain the coupling constants. For example, for 2 electrons, $J$ can be obtained from the energy difference between the lowest singlet state (ground state) and lowest triplet state (first excited state).
%
%
\section{Results}
\label{results}
\subsection{Electronic density}
\label{charge_density}
To verify that the electrons are indeed localized in the 1D Wigner crystals, we analyze the charge distribution of the electrons in their ground state. 
To this aim, we calculated the density distribution using the approach described in Sec. \ref{dens_distr} for Wigner crystals containing two, three and four electrons.
In practice, we have to truncate the expansion in Eq.~\eqref{MBeigfs} for the calculation of the ground-state wave function $\Psi_0$.
In our calculations of the density, we consider all Slater determinants defined up to a maximum integer number $n$ which corresponds to the index of the single-particle basis functions in Eq.~\eqref{eigf}. 
In the case of $N=2$ we used $n=63$ which corresponds to 7875 Slater determinants, while for $N=3$ and $N=4$ we used $n=29$ and $n=17$, respectively, corresponding to 30856 and 46376 Slater determinants. At these basis sizes, the density profiles are converged. We refer to Appendix \ref{dens_conv} for the convergence analysis.

\begin{figure}[t]
    \centering
    \begin{subfigure}[t]{0.47\textwidth}
        \centering
        \includegraphics[width=\linewidth]{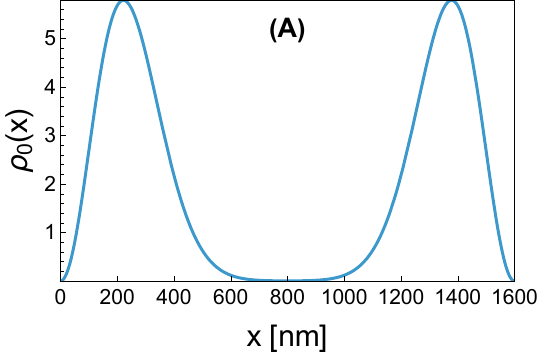}
        \label{densities2el}
    \end{subfigure}
    \hfill
    \begin{subfigure}[t]{0.47\textwidth}
        \centering
        \includegraphics[width=\linewidth]{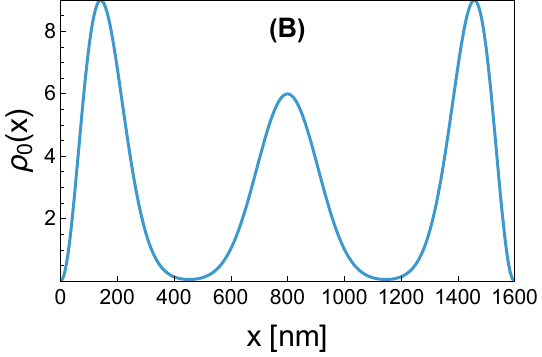}
        \label{densities3el}
    \end{subfigure}

    \vspace{0.6cm}

    \begin{subfigure}[t]{0.47\textwidth}
        \centering
        \includegraphics[width=\linewidth]{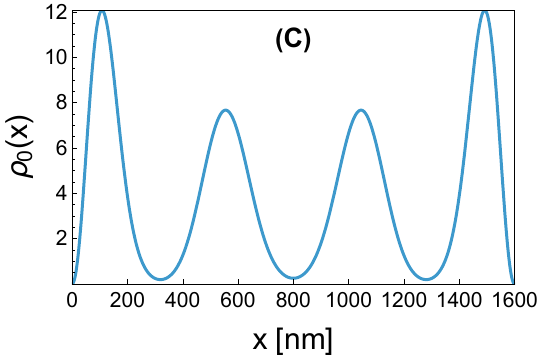}
        \label{densities4el}
    \end{subfigure}

\caption{Ground-state electronic density distributions for 2, 3, and 4 electrons confined to a  nanotube with a length of 1600 nm; \textbf{(A)} : two electrons \textbf{(B)} three  electrons. \textbf{(C)} four electrons.}
\label{fig:densities}
\end{figure}

In Fig. \ref{fig:densities} we report the density distribution of 1D Wigner crystals system with two, three and four electrons 
confined to a one-dimensional box of length $L=1600 \, \mathrm{nm}$.
We clearly observe the localization of the electrons in the nanotube from the number of peaks in each spectrum with each peak being separated from its neighbor by a deep valley in which the electron density nearly vanishes. 
We also observe that the peaks at the border of the system are bigger and sharper than the central peaks. 
This is due to the hard walls we imposed at the boundaries and the very strong repulsion between the electrons.
Because of the strong repulsion the outermost electrons are pushed up against the hard walls.

\begin{figure}[t]
	\centering
	\includegraphics[width=\columnwidth]{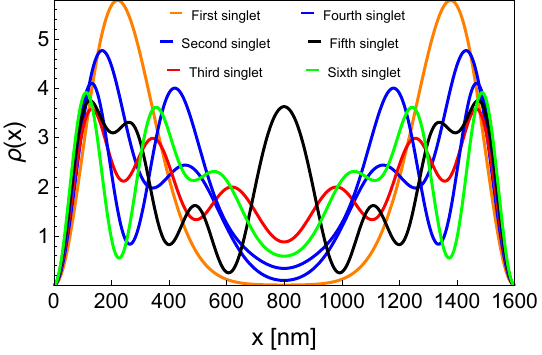}
	\caption{Density profiles corresponding to the ground state and the first five excited states in the singlet sector of the Wigner dimer in a nanotube of length $L=1600 \, \mathrm{nm}$. The density profiles of the triplet sector are almost identical.}
	\label{excitedstates}
\end{figure}

We note that in the Wigner regime we study here, the spin symmetry of the wave function has no influence on the density distribution~\cite{FCI_gauss,Escobar2024}.
This is because, at strong interaction, the electrons behave as a single particle, and
fixing the position of one electron uniquely determines the positions of the other electrons.
In Fig.~\ref{excitedstates} we report the density distributions of the first five excited states together with the ground-state density profile of the Wigner dimer for the singlet sector. The density profiles corresponding to the the triplet sector are almost identical.
We observe the emergence of peaks in the central region of the domain in the density distributions of the excited states. This suggests that, as expected, for the excited states, the electrons are, on average, closer to each other than in the ground state. In particular, it is interesting to see that the density profile corresponding to the fourth excited state exhibits a pronounced peak in the middle of the domain.

\begin{figure}[b]
	\centering
	\includegraphics[width=\columnwidth]{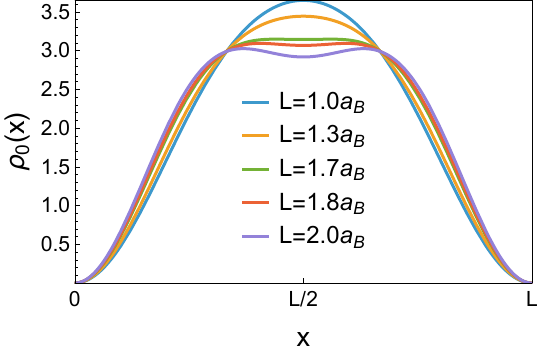}
	\caption{Ground-state density profiles of the Wigner dimer for various values of the length of the nanotube.}
	\label{crossover}
\end{figure}

Finally, we study the crossover from the Fermi-liquid regime to the Wigner regime for the case of two electrons. At high electronic density (small $L$), in the Fermi-liquid regime, we expect the density profile corresponding to the ground-state singlet wave function to exhibit a single maximum at the center of the domain, since in accordance with the Pauli principle, electrons with opposite spin are allowed to occupy the same position. 
Instead, at low electron density (large $L$), in the Wigner regime, the density profile corresponding to the ground-state singlet wave function exhibits two maxima as shown in panel (A) of Fig.~\ref{fig:densities}.
Therefore, we can estimate the crossover from the Fermi-liquid regime to the Wigner regime from the number of maxima.
In Fig.~\ref{crossover} we report the ground-state density distributions for various values of the length $L$ of the box.
We observe that number of maxima in the density distribution goes from 1 to 2  around $L\approx 1.7 a_B$. We conclude that the crossover point between the two regimes lies around this value for the length of the box.
\subsection{The pair density}
\begin{figure}[t]
	\centering
	\includegraphics[width=\columnwidth]{"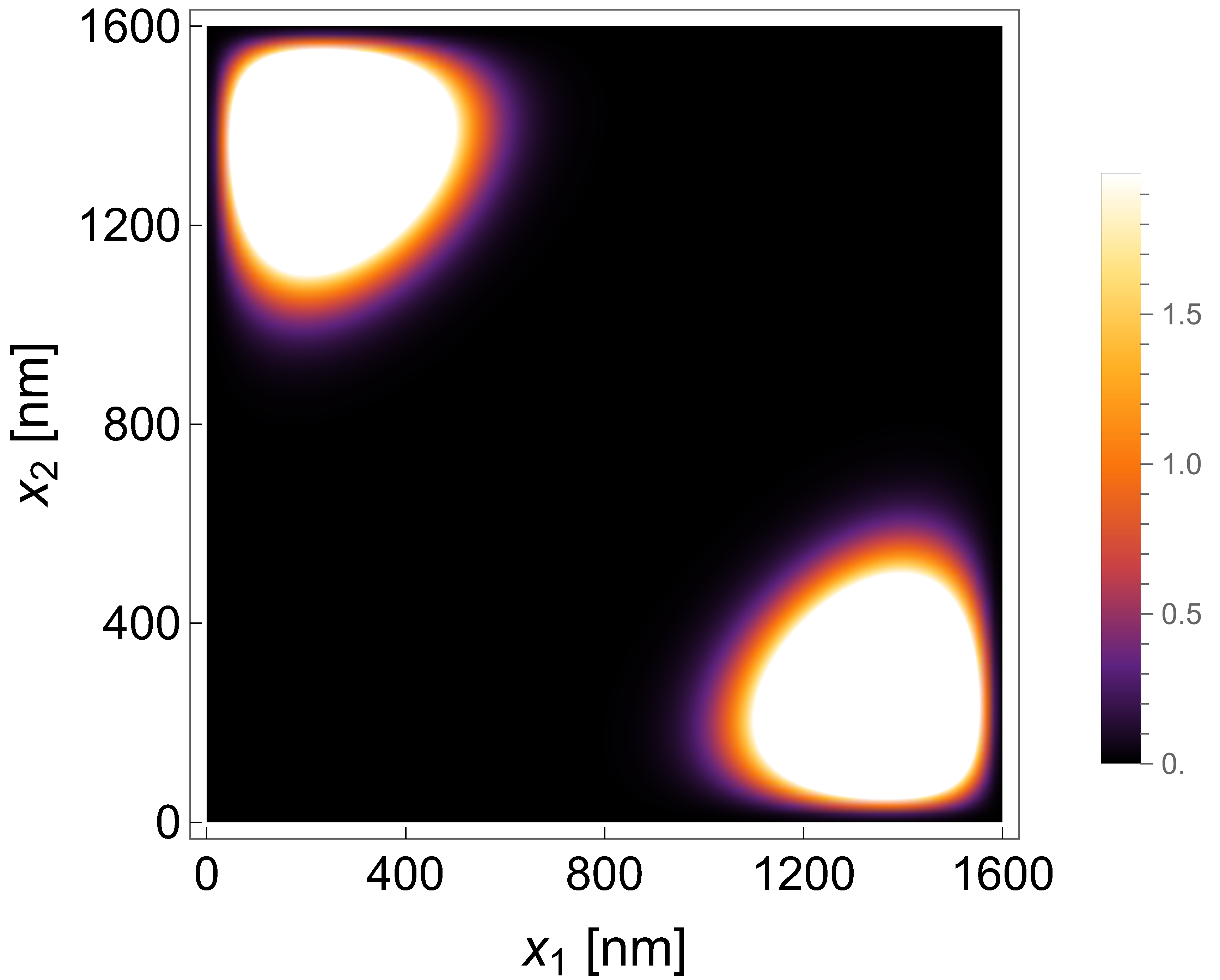"}
	\caption{The pair density $\Gamma(x_1,x_2)$ corresponding to the ground-state wave functions of a Wigner dimer in a nanotube of length $L=1600 \, \mathrm{nm}$.}
	\label{Gamma2el}
\end{figure}

The electronic density could exhibit Friedel-like oscillations which could potentially weaken the conclusion that the electrons are localized.
Therefore, in the case of the Wigner dimer, we also calculated its pair density $\Gamma(x_1,x_2)$ which gives the joint probability of finding the electrons at positions $x_1$ and $x_2$.
In Fig.~\ref{Gamma2el} we report the pair density for the Wigner dimer in a box of length $L=1600 \, \mathrm{nm}$ corresponding to the length of the nanotube in the experiment.
We clearly observe a large probability of finding the two electrons on opposite sides of the box. Instead, finding the two electrons close to each other has almost zero probability.
This clearly demonstrates that the electrons are indeed localized.
\subsection{The exchange coupling constant}
\label{exchange_coup}
In the previous section we showed from the density distributions that the electrons in the quantum are truly localised.
Therefore, we can model them with the HDVV model discussed in Sec.~\ref{ab_initio_calc}.
In this work we will illustrate our approach by applying it to the case of 2 electrons.
The corresponding HDVV Hamiltonian is given by
\begin{equation}
	\hat{H}_{HDVV}=J\hat{S}_1\cdot\hat{S}_2\,.
	\label{Heis_2el}
\end{equation}
It is convenient to represent $\hat{H}_{HDVV}$ in the spin basis given by $\ket{\uparrow \uparrow}, \ket{\uparrow \downarrow}, \ket{\downarrow \uparrow}, \ket{\downarrow \downarrow}$.
We thus obtain
\begin{equation}
   	\hat{H}_{HDVV}=J
   	\begin{pmatrix}
   		\frac{\hbar^2}{4} & 0 & 0 & 0 \\
   		0 & -\frac{\hbar^2}{4} & \frac{\hbar^2}{2} & 0\\
   		0 & \frac{\hbar^2}{2} & -\frac{\hbar^2}{4} & 0\\
   		0 & 0 & 0 & \frac{\hbar^2}{4}
   	\end{pmatrix}.
\end{equation}
Diagonalizing the above matrix yields four eigenvectors, three of which are degenerate.
The eigenvector thus correspond to a singlet (S) and a triplet state (T).
Their eigenenergies $E_S$ and $E_T$ are
\begin{equation}
    	E_S=-\frac{3\hbar^2}{4}J\, , \qquad E_T=\frac{\hbar^2}{4}J\,,
\end{equation}
and the singlet-triplet splitting is thus equal to the exchange constant $J$, i.e., 
\begin{equation}
    	J=(E_T-E_S)/\hbar^2\,.
        \label{relation_J2}
\end{equation}
We can thus obtain a value for $J$ by calculating the singlet-triplet splitting in the \textit{ab initio} calculation of the 1D Wigner crystal with 2 electrons.
This corresponds to the difference between the energy of the ground state (a singlet) and the energy of the first excited state (a triplet).
As explained in Sec.~\ref{spin_adaptation}, in order to avoid the unphysical mixing of spin states in the calculation, we use the spin adapted basis to calculate the lowest singlet and triplet energies separately.

\begin{figure}[b]
	\centering
	\includegraphics[width=\columnwidth]{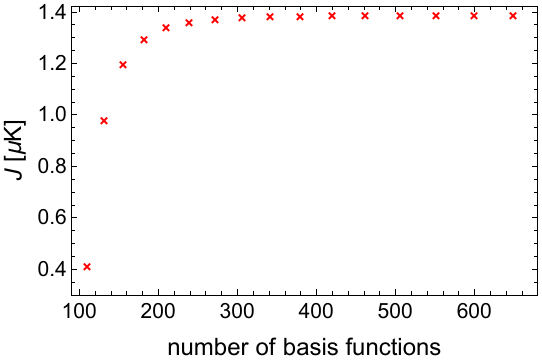}
	\caption{The exchange coupling constant $J$ as a function of the number of basis functions for a system of two electrons confined to a nanotube of length $L=1600\, \mathrm{nm}$. }
	\label{plot_relationsJ_2}
\end{figure}

As for the electronic density, in calculating $J$ we consider all Slater determinants defined up to a maximum integer number $n$ which corresponds to the index of the single-particle basis functions in Eq.~\eqref{eigf}. 
In Fig.~\ref{plot_relationsJ_2} we show the convergence of the exchange coupling constant $J$ as a function of $n$.
We see that the value of $J$ converges as a function of $n$ and at convergence its value is given by 
\begin{equation}
    	J= 1.39  \, \mu \mathrm{K}\,.
    	\label{J_2el}
\end{equation}
The value for $J$ we obtain is similar to that found in Ref.~\cite{Experiment}, where they obtained $J=1.9 \, \mu \mathrm{K}$ using the density-matrix renormalisation group. The discrepancy in the results can be due to the Hamiltonian used in Ref.~\cite{Experiment} which has a quartic potential to contain the electrons in addition to the hard walls we impose here.
However, the addition of the same quartic potential as in Ref.~\cite{Experiment} to the Hamiltonian in Eq.~\eqref{Ham} does not induce a large modification of the exchange coupling. In fact, it slightly reduces the exchange coupling to $J = 1.01\, \mu \mathrm{K}$ which means that it is not the origin of the difference. The details of the calculation with the quartic potential are given in appendix \ref{quartic_pot}. 
Another source of discrepancy arises from the fact that an effective SU(4) symmetrical exchange Hamiltonian model, originally derived in Ref.~\cite{pot_rin}, is used to extract $J$ which is different from the HDVV Hamiltonian in Eq.~\eqref{eff_Ham} that we use.

\begin{figure}[t]
	\centering
	\includegraphics[width=\columnwidth]{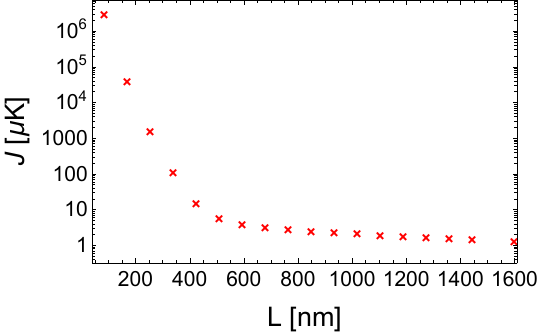}
	\caption{The exchange coupling constant $J$ as a function of the length $L$ of the nanotube.}
	\label{J_various_L}
\end{figure}

We also investigated the behaviour of the exchange coupling $J$ as a function of the length of the carbon nanotube. We report the results in Fig.~\ref{J_various_L}. We see that the exchange coupling decreases monotonously as a function of $L$ and is expected to vanish in the limit $L\rightarrow\infty$.

Finally, we note that the numerical approach presented here can be extended to compute the exchange coupling constant of systems with more than two electrons. 
Indeed, we observe from the charge distribution in Fig. \ref{fig:densities} that the peaks are evenly distributed over the domain, so we expect that the Heisenberg-Dirac-Van Vleck Hamiltonian in Eq. \eqref{eff_Ham} correctly represents the low-energy properties of these Wigner molecules.
%
\section{Conclusions}
\label{conclusion}
We developed an approach to study one-dimensional Wigner crystals with few electrons from first principles.
In particular, we simulated an experimental setup in which electrons were injected in a carbon nanotube of length $L=1600$ nm by confining electrons to a nanotube of the same size with hard walls.
We used full configuration interaction in order to capture all the electron correlation for a given basis set.
We calculated the electronic density distribution from which we clearly observed the localisation of the electrons in the Wigner crystals. 
Finally, we also calculated the exchange coupling constant of the Wigner dimer and we found a good agreement with a value in the literature.

As future developments of this work, we plan to compute the exchange coupling constants for systems with more than two electrons. To avoid unphysical mixing of different spin-symmetry states we have to create spin adapted basis functions similar to the ones we created for two electrons.
Finally, we note that full-configuration interaction, which scales exponentially with the number of electrons, becomes prohibitive for one-dimensional Wigner molecules with many electrons.
Alternative methods to full-configuration interaction that are numerically more efficient for studying
large Wigner molecules are the CASSCF (Complete Active Space Self-Consistent Field) method~\cite{Roos_1980} and the DMRG-SCF method~\cite{Zgid_2008}. Our numerically exact FCI results provide a benchmark for these approximate methods.

\section{Acknowledgments}
This project has received financial support from the CNRS through the MITI interdisciplinary programs.
JAB thanks the French “Agence Nationale de la Recherche (ANR)” for financial support (grant agreement no.
ANR-22-CE29-0001).

\appendix
\section{Optimized basis}
\label{Optimized_basis}
The Hamiltonian in Eq.~\eqref{adham} has inversion symmetry with respect to $x=0.5$ and the basis functions $\chi_n$ in Eq.~\eqref{eigf} are either symmetric (if $n$ is odd) or anti-symmetric (if $n$ is even) with respect to inversion.
Therefore, the matrix elements $A[i,j,k,l]$ given in Eq.~\eqref{Aijkl1} have the following properties
\begin{equation}
    A[i,j,k,l]=
    \begin{cases}
    0 \qquad \text{if $i+j$ is even and $k+l$ is odd}\\
    0 \qquad \text{if $i+j$ is odd and $k+l$ is even}
    \end{cases}.
    \label{propertyAijkl}
\end{equation}
%
From the general expressions of the matrix elements in the singlet and triplet sectors, defined in Eq.~\eqref{singlet_matel} and \eqref{triplet_matel}, we observe that, due to property in Eq. \eqref{propertyAijkl}, the basis wavefunctions with indices such that $i+j=$ even are orthogonal to those that have indices $k+l=$ odd. 
From these relations we conclude that the ground-state wave function, which is the lowest singlet eigenfunction of $\hat{H}$, is written in terms of spin-adapted singlet functions $\Psi^{S}_{m,n}$ that satisfy the condition that $m+n$ has to be even.
If, instead, we consider the lowest triplet state, due to the relation in Eq. \eqref{propertyAijkl}, it is expanded in spin-adapted triplet functions $\Psi^{T}_{m,n}$ that satisfy the condition that $m+n$ has to be odd. 

\begin{figure}[t]
	\centering
	\includegraphics[scale=0.9]{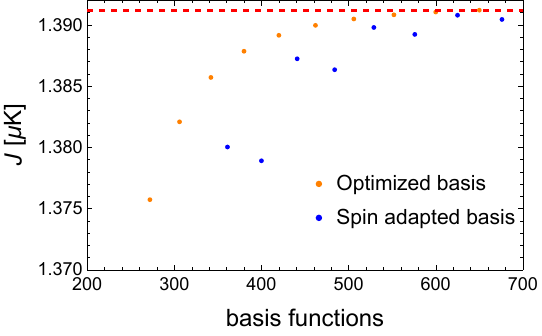}
	\caption{Convergence of the exchange coupling $J$ for a system of two electrons using the standard spin-adapted basis (blue dots) and the optimized spin-adapted basis (orange dots) at $L=1600\, \mathrm{nm}$. 
    The red dashed line represents the converged value of $J$ obtained with the optimized spin adapted basis and taking into account 650 Slater determinants. }
	\label{basis_2methods}
\end{figure}

From the diagonalization of $\hat{H}$ in Eq.~\eqref{adham}, we identified an empirical sorting criterion for the singlet and triplet basis wavefunctions in systems with length $L = 1600 \, \mathrm{nm}$. This criterion allows us to accelerate the convergence of either the exchange coupling or the density profiles. 
In particular, we observed that the singlet and triplet basis functions contributing most significantly to the ground state and the first excited state, respectively, are those characterized by the lowest values of the sum of their indices.
Therefore, in practice, in our calculations we include all Slater determinants of which the sum of the indices of the singlet or triplet basis functions is equal or smaller than a given integer.
%
In Fig. \ref{basis_2methods}  we show, at  $L = 1600 \, \mathrm{nm}$, the convergence of the exchange coupling $J$ using the optimized spin-adapted basis as a function of the number of basis functions. In the figure, we compare it with the convergence of $J$ computed with the standard spin-adapted basis, defined in Sec.~\ref{spin_adaptation}. The exchange couplings are computed with different basis functions. We observe that the two basis choices lead to the same converged value of $J$, but the optimized basis allows to reach faster the numerical convergence than the spin-adapted basis.
\FloatBarrier
\section{Convergence of the density distributions}
\label{dens_conv}
\begin{figure}[t]
	\centering
	\includegraphics[width=\columnwidth]{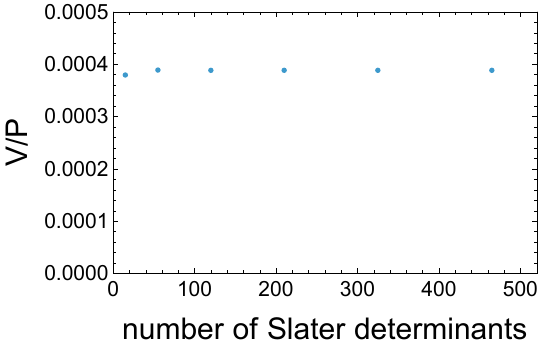}
	\caption{Valley-to-peak ratio (V/P) for a Wigner dimer as a function of the number of Slater determinants.}
	\label{conv_dens2el}
\end{figure}

We study the convergence of the ground-state density distributions of the Wigner systems with 2, 3 and 4 electrons with respect to the size of the basis set.
The density profile of the Wigner dimer shows one valley and two peaks that are equivalent by symmetry.
Therefore, in order to demonstrate the convergence of the density profile, we report the valley-to-peak ratio as a function of the number of Slater determinants in Fig.~\ref{conv_dens2el}.
We observe that several hundreds of Slater determinants are sufficient to converge the valley-to-peak ratio which is well below the 7875 Slater determinants used to calculate the density profile of the Wigner dimer in Fig.~\ref{fig:densities}.
%

\begin{figure}[b]
	\centering
	\includegraphics[width=\columnwidth]{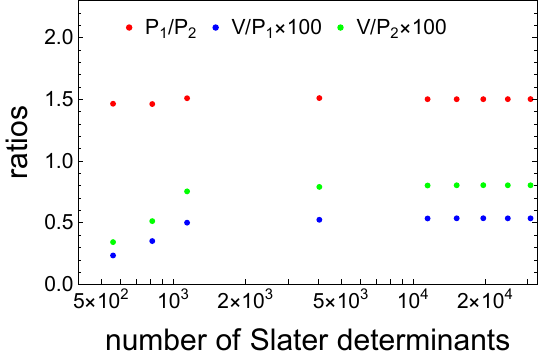}
	\caption{Peak-to-peak ratio (P$_1$/P$_2$) of an outermost peak (P$_1$) and the central peak (P$_2$) and valley to peak ratios (V/P$_1$ and V/P$_2$) as a function of the number of Slater determinants.}
	\label{conv_dens3el}
\end{figure}

For the case of three electrons there are three peaks, of which the two outermost are equivalent by symmetry, and two valleys that are also equivalent by symmetry.
In Fig.~\ref{conv_dens3el} we report the peak-to-peak ratio as well as the two valley-to-peak ratios.
We see that for this case a few thousand Slater determinants are required to reach convergence.

\begin{figure}[t]
	\centering
	\includegraphics[width=\columnwidth]{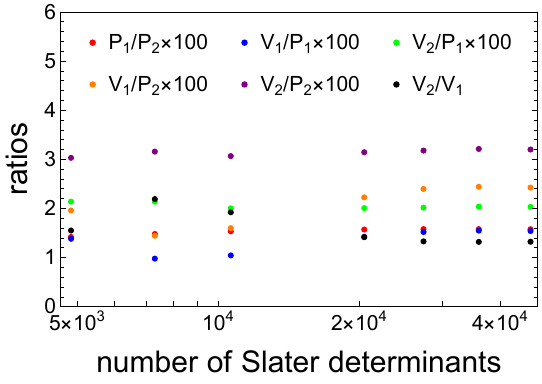}
	\caption{Peak-to-peak ratios (P$_1$/P$_2$) of an outermost peak (P$_1$) and an innermost peak (P$_2$), valley-to-valley ratios (V$_2$/V$_1$) of an innermost valley (V$_2$) and an outermost valley (V$_1$) and valley-to-peak ratios (V$_1$/P$_1$, V$_2$/P$_2$, V$_1$/P$_2$ and V$_2$/P$_1$) as a function of the number of Slater determinants.}
	\label{conv_dens4el}
\end{figure}

Finally, for the case of four electrons there are 4 peaks, of which the two outermost and the two innermost peaks are equivalent by symmetry, and three valleys, of which the two outermost are equivalent by symmetry. 
In Fig.~\ref{conv_dens4el} we report all the peak-to-peak, valley-to-peak and valley-to-valley ratios. We observe that for four electrons we need of the order of 10 000 Slater determinants to reach convergence.
\section{The Wigner dimer in a confining quartic potential}
\label{quartic_pot}
In this section we verify the influence of adding the quartic potential used in Ref.~\cite{Experiment} to the Hamiltonian in Eq.~\eqref{Ham} on the exchange coupling constant of the Wigner dimer confined to the nanotube.
The Hamiltonian including a quartic potential centered in the middle of the domain,
\begin{align}
    \hat{H}_Q=-\frac{\hbar^2}{2m_{\text{eff}}}\sum_{i}^{N}\frac{\partial^2}{\partial z^2_i}+&\frac{e^2}{2}\sum_{i\not = j}V(z_i-z_j)+ \notag \\
    &\frac{1}{4}A\sum_{i=1}^{N}\bigg(z_i-\frac{L}{2}\bigg)^4\,, 
    \quad z_i\in[0,L]\,.
    \label{Ham_quartic}
\end{align}
where the parameter $A$ controls the width of the confining potential.
As we did before, we rescale the coordinates according to $x_i=z_i/L$ and obtain
\begin{align}
    \hat{H}_Q=\frac{\hbar^2}{2m_{\text{eff}}L^2}\sum_{i}^{N}&\frac{\partial^2}{\partial x^2_i}+\frac{2L}{a_B}\sum_{i\not = j}V(x_i-x_j)+ \notag \\
    &\frac{1}{2}\bigg(\frac{L}{l_d}\bigg)^{6}\sum_{i=1}^{N}\bigg(x_i-\frac{1}{2}\bigg)^4\,, 
    \quad x_i\in[0,1]\,.
\end{align}
where $l_d=\bigg(\frac{\hbar^2}{m_{\text{eff}}A}\bigg)^{1/6}$.

\begin{figure}[h]
	\centering
	\includegraphics[width=\columnwidth]{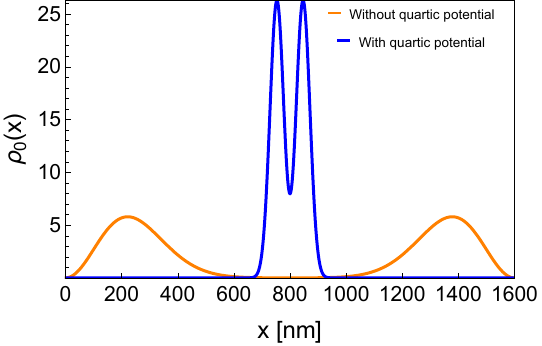}
	\caption{Ground-state density distributions of a system of two electrons in a nanotube with and without a quartic potential in the Hamiltonian.}
	\label{densities_quartic}
\end{figure}

In the singlet sector, the matrix elements associated to the quartic potential are written as
\begin{align}
    & \bra{\Psi^S_{m,n}}\Big(x_1-\frac{1}{2}\Big)^4+\Big(x_2-\frac{1}{2}\Big)^4\ket{\Psi^S_{p,q}} = \notag \\
    &\frac{1}{\sqrt{(1+\delta_{m,n})(1+\delta_{p,q})}} \notag\\
    &\Bigg[\delta_{n,q}\int_0^1\chi_m(x)\Big(x-\frac{1}{2}\Big)^4\chi_p(x)\, dx +\notag\\
    &\delta_{n,p}\int_0^1\chi_m(x)\Big(x-\frac{1}{2}\Big)^4\chi_q(x) \, dx + \notag\\
    &\delta_{m,q}\int_0^1 \chi_n(x)\Big(x-\frac{1}{2}\Big)^4\chi_p(x)\,dx + \notag\\
    &\delta_{m,p}\int_0^1 \chi_n(x)\Big(x-\frac{1}{2}\Big)^4\chi_q(x)\,dx \Bigg]\,.
    \label{singlet_matelQ}
\end{align}
If we consider the contribution of the quartic potential to the matrix elements of the triplet sector, we find that 
\begin{align}
    & \bra{\Psi^T_{m,n}}\Big(x_1-\frac{1}{2}\Big)^4+\Big(x_2-\frac{1}{2}\Big)^4\ket{\Psi^T_{p,q}} = \notag \\
    &\delta_{n,q}\int_0^1\chi_m(x)\Big(x-\frac{1}{2}\Big)^4\chi_p(x)\, dx -\notag\\
    &\delta_{n,p}\int_0^1\chi_m(x)\Big(x-\frac{1}{2}\Big)^4\chi_q(x) \, dx - \notag\\
    &\delta_{m,q}\int_0^1 \chi_n(x)\Big(x-\frac{1}{2}\Big)^4\chi_p(x)\,dx + \notag\\
    &\delta_{m,p}\int_0^1 \chi_n(x)\Big(x-\frac{1}{2}\Big)^4\chi_q(x)\,dx \,.
    \label{triplet_matelQ}
\end{align}
Therefore, the general expression of the matrix elements of the singlet sector is
\begin{align}
    &\bra{\Psi^S_{m,n}}\hat{H}_Q\ket{\Psi^S_{p,q}} = \bra{\Psi^S_{m,n}}\hat{H}\ket{\Psi^S_{p,q}} +  \\
    &\frac{1}{2}\bigg(\frac{L}{l_d}\bigg)^{6} \bra{\Psi^S_{m,n}}\Big(x_1-\frac{1}{2}\Big)^4+\Big(x_2-\frac{1}{2}\Big)^4\ket{\Psi^S_{p,q}} \,.
\end{align}
For the triplet sector, the general expression of Hamiltonian is
\begin{align}
    &\bra{\Psi^T_{m,n}}\hat{H}_Q\ket{\Psi^T_{p,q}} = \bra{\Psi^T_{m,n}}\hat{H}\ket{\Psi^T_{p,q}} + \\
    &\frac{1}{2}\bigg(\frac{L}{l_d}\bigg)^{6} \bra{\Psi^T_{m,n}}\Big(x_1-\frac{1}{2}\Big)^4+\Big(x_2-\frac{1}{2}\Big)^4\ket{\Psi^T_{p,q}} \,.
\end{align}
In Ref.~\cite{Experiment} $l_d$ was chosen to be equal to one tenth of the length of the nanotube, i.e., $l_d = 160 \, \text{nm}$. This is also the value we used in our calculation. 
In Fig. \ref{densities_quartic} we compare the ground-state density profiles obtained with and without the quartic potential. 
We observe that, as expected, the confining potential pushes the electrons towards the center of the well.

\clearpage

\nocite{*}
\bibliographystyle{unsrt} 
\bibliography{references} 

\end{document}